\documentclass[12pt,preprint]{aastex}
\usepackage{graphicx}
\usepackage{epsfig}
\usepackage{multirow}
\usepackage{footnote}

\shorttitle{Optical and Infrared Study of GALEX1931}
\shortauthors{C. Melis et al.}

\begin{document}


\title{Accretion of a Terrestrial-Like Minor Planet by a White Dwarf}


\author{Carl Melis\altaffilmark{1,6}, J. Farihi\altaffilmark{2}, P. Dufour\altaffilmark{3}, B. Zuckerman\altaffilmark{4}, Adam J. Burgasser\altaffilmark{1,5}, P. Bergeron\altaffilmark{3}, J. Bochanski\altaffilmark{5}, R. Simcoe\altaffilmark{5}}
\email{cmelis@ucsd.edu}


\altaffiltext{1}{Center for Astrophysics and Space Sciences, University of California, San Diego, CA 92093-0424, USA}
\altaffiltext{2}{Department of Physics \& Astronomy, University of Leicester, Leicester LE1 7RH, UK}
\altaffiltext{3}{D\'{e}partement de Physique, Universit\'{e} de Montr\'{e}al, Montr\'{e}al, QC H3C 3J7, Canada}
\altaffiltext{4}{Department of Physics and Astronomy, University of California,
Los Angeles, CA 90095-1547, USA}
\altaffiltext{5}{Massachusetts Institute of Technology, Kavli Institute for Astrophysics and Space Research, Building 37, Room 664B, 77 Massachusetts Avenue, Cambridge, MA 02139, USA}
\altaffiltext{6}{Joint CASS Departmental Fellow and NSF AAPF Fellow}


\begin{abstract}
We present optical and infrared characterization of the polluted
DAZ white dwarf GALEX\,J193156.8+011745.
Imaging and spectroscopy
from the ultraviolet to the thermal infrared indicates that
the white dwarf hosts excess infrared emission consistent
with the presence of an orbiting dusty debris disk. 
In addition to the five elements previously identified,
our optical echelle spectroscopy reveals chromium and manganese
and enables restrictive upper limits on several other elements.
Synthesis of all detections and upper limits suggests that the white dwarf 
has accreted a differentiated parent body.
We compare the inferred bulk elemental composition of the accreted
parent body to expectations for the bulk composition of 
an Earth-like planet stripped
of its crust and mantle and find relatively good agreement. 
At least two processes could be important in shaping
the final bulk elemental composition of rocky bodies during the
late phases of stellar evolution: irradiation and interaction
with the dense stellar wind.
\end{abstract}





\section{Introduction}

Recent studies have solidified the paradigm that metal-rich white dwarfs are
polluted by remnant rocky bodies from their planetary systems
\citep[e.g.,][and references therein]{zuckerman07,jura08,farihi09,farihi10b,farihi10a,dufour10,klein10,melis10,zuckerman10}.
It is now desirable to understand the mass, nature, and history of the
accreted parent body at each white dwarf. The larger the degree
of pollution, the greater potential for insight into extrasolar rocky bodies.

\citet{vennes10} describe a heavily polluted DAZ white dwarf
$-$ GALEX\,J193156.8+011745 (hereafter GALEX1931) $-$ exhibiting atmospheric
magnesium, silicon, oxygen, calcium, and iron (Mg, Si, O, Ca, and Fe respectively).
Additionally, Two Micron All Sky Survey \citep[2MASS][]{skrutskie06} 
photometry suggested excess near-infrared
emission. Abundance analysis and the shape of the putative infrared excess
was interpreted by \citet{vennes10} as consistent with a mid L-type dwarf 
that orbits and pollutes the white dwarf. However, near-infrared
emission from a prominent
dust disk is also consistent with the spectral energy distribution
presented by \citet{vennes10}. \citet{debes10}
report Wide-field Infrared Survey Explorer (WISE) 
imaging and optical spectroscopic monitoring of 
GALEX1931 that support a disk interpretation. 

Here we present optical through infrared imaging and spectroscopy
of GALEX1931 designed to better constrain the presence of infrared
excess emission and to probe the material polluting the stellar
atmosphere.


\section{Observations}
\label{secobs}

\subsection{Nickel Optical Imaging}

Optical imaging was performed on UT 25 May 2010 at Lick
Observatory with the 40" Nickel telescope. These observations
used the facility's Direct Imaging Camera (CCD-C2), a
2048 $\times$ 2048 pixel detector with 15 $\mu$m pixels. The
0.184$^{\prime\prime}$ pixel$^{-1}$ plate scale affords a field
of view of $\sim$6.3$^{\prime}$ squared. The detector was
binned by two in rows and columns and was readout in fast mode.
Occasional clouds passed
through the region of the sky being observed; frames affected
by clouds were not used in the final analysis. 

GALEX1931 was observed in each of the BVRI filters
\citep{bessell90}. For each filter a 4-step dither pattern with 
10$^{\prime\prime}$ steps was repeated with 
20 second integrations per step position. 
The flux calibrator star SA 111-1925 \citep{landolt09} was
visited following the completion of a dither sequence for each filter. 
Images are reduced by median-combining all
good frames to obtain a ``sky'' frame and subtracting
this sky frame from each image. Sky-subtracted images are then 
divided by flat field frames obtained by imaging the twilight sky
(BVR) or the illuminated telescope dome (I). Each science frame is
registered using bright stars in the field and then all science 
frames are median combined to yield the final reduced image.
Detector counts for GALEX1931 and SA 111-1925 
are extracted with an aperture that yields
$\sim$85\% encircled energy (with a negligible correction between
the two stars). This is achieved by extracting counts for both
stars with a 6 pixel (2.2$^{\prime\prime}$) radius circular aperture.
Sky is sampled with an annulus extending from 20-100 pixels.
Uncertainties are derived from the dispersion of measurements
made from individually reduced frames. The uncertainties for
SA 111-1925 are propagated into the final quoted uncertainty
for GALEX1931.
The final V-band measurement from the Nickel observations is
less accurate than that reported in \citet{vennes10}; we thus
adopt the \citet{vennes10} V-band measurement.

\begin{figure}
\begin{minipage}[t!]{40mm}
 \includegraphics[width=40mm]{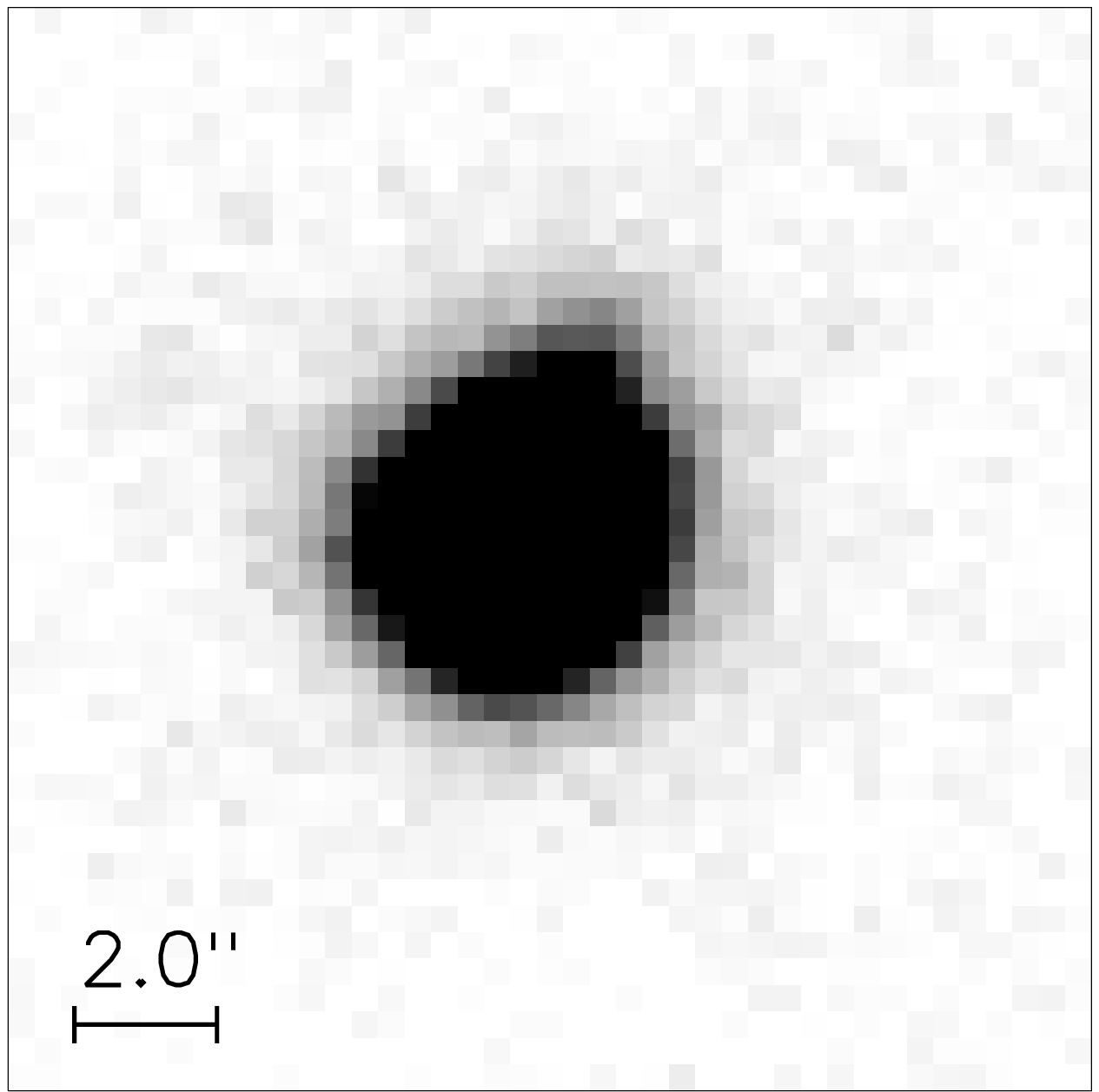}
\end{minipage}
\begin{minipage}[t!]{40mm}
 \includegraphics[width=40mm]{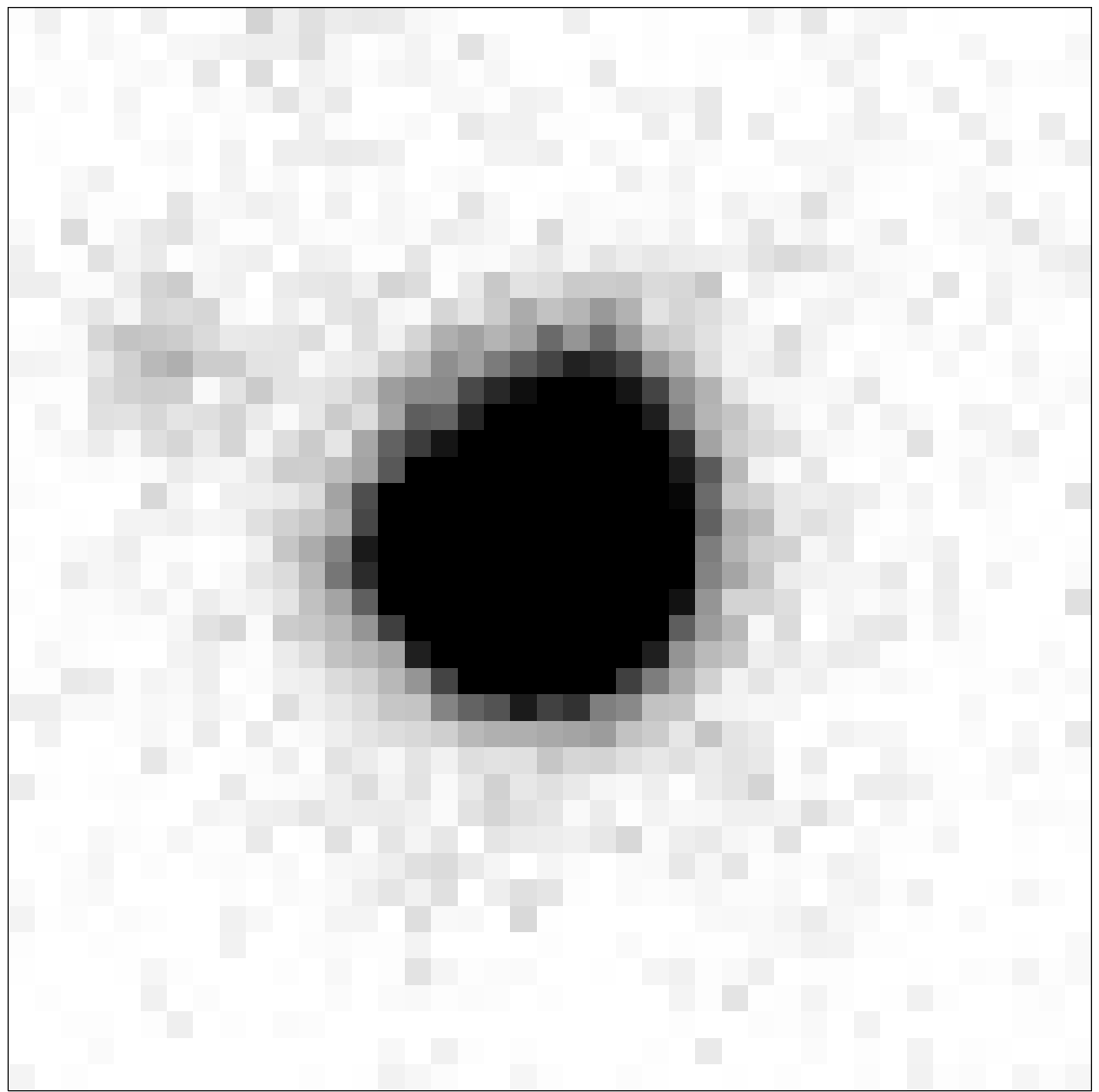}
\end{minipage}
\begin{minipage}[t!]{40mm}
 \includegraphics[width=40mm]{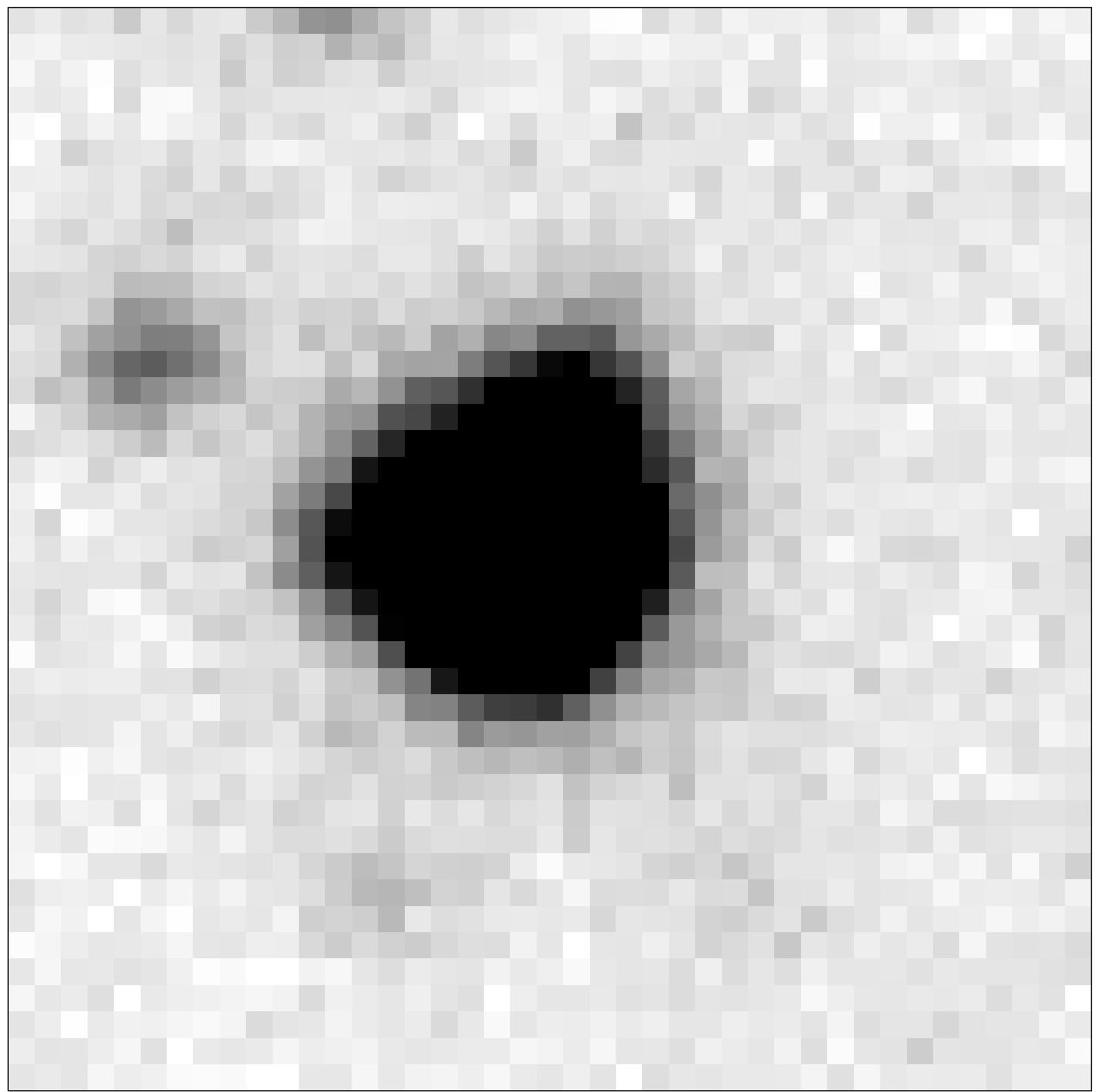}
\end{minipage}
\begin{minipage}[t!]{40mm}
 \includegraphics[width=40mm]{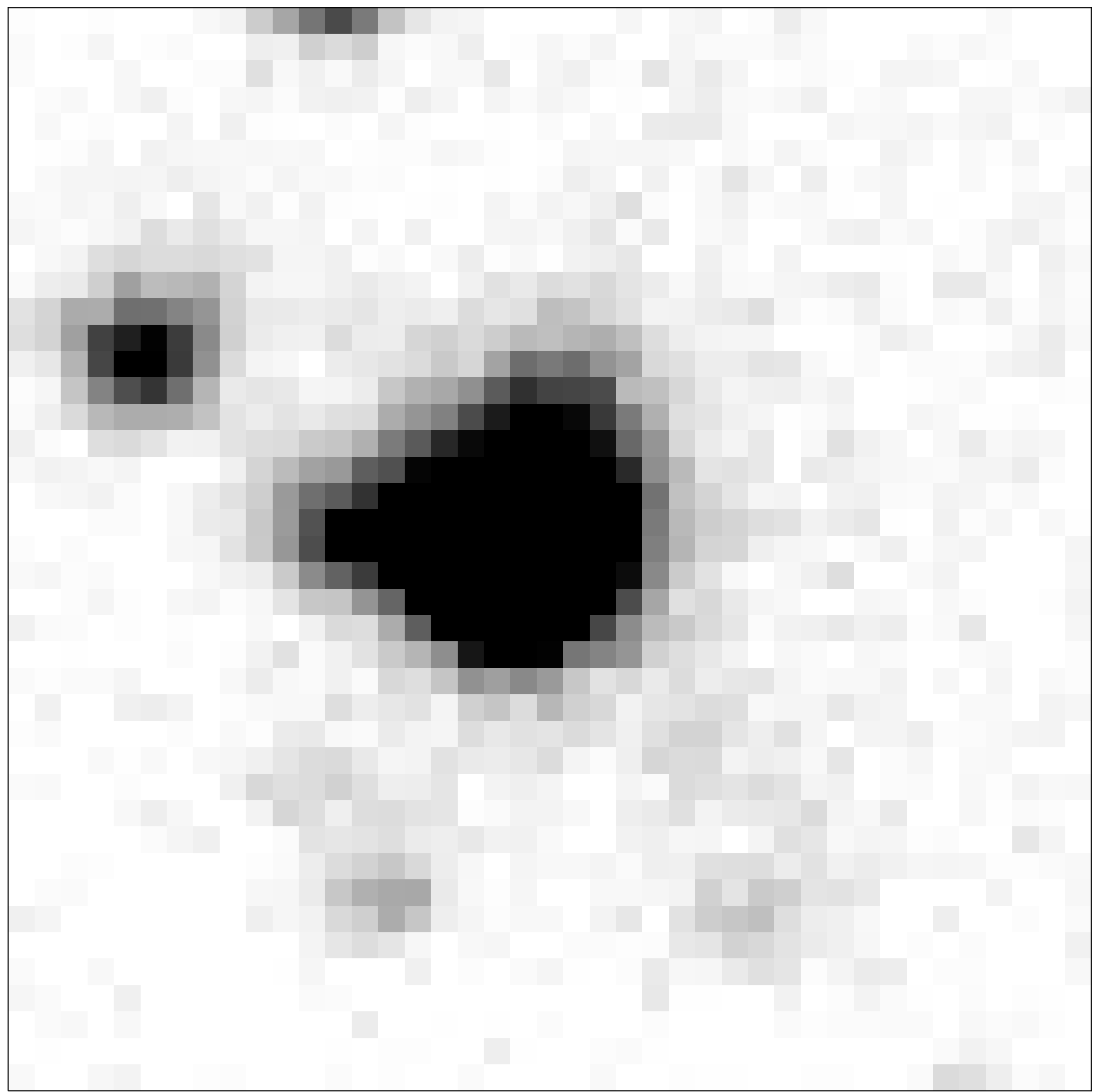}
\end{minipage}
\\*
\begin{minipage}[t!]{40mm}
 \includegraphics[width=40mm]{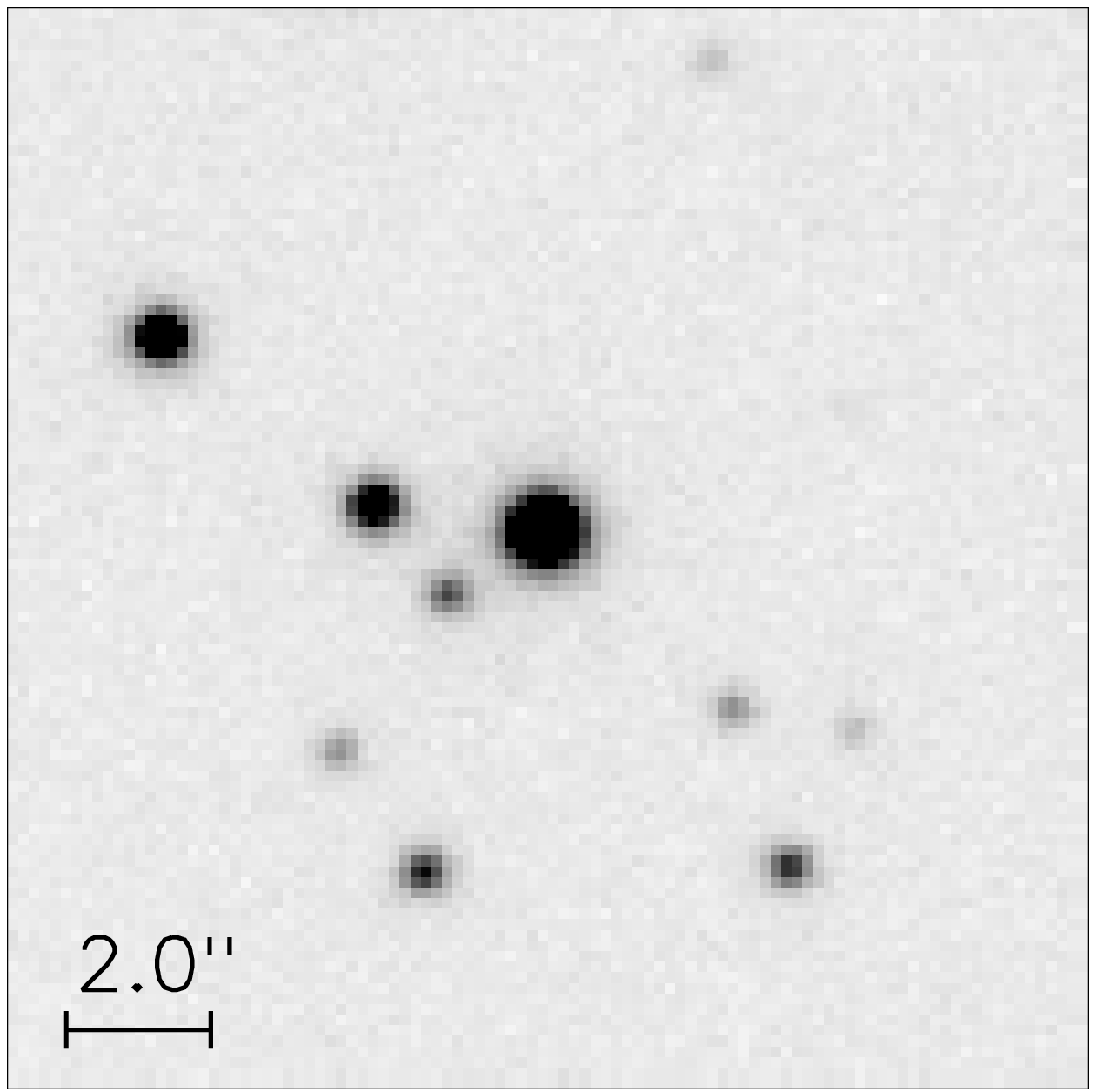}
\end{minipage}
\begin{minipage}[t!]{40mm}
 \includegraphics[width=40mm]{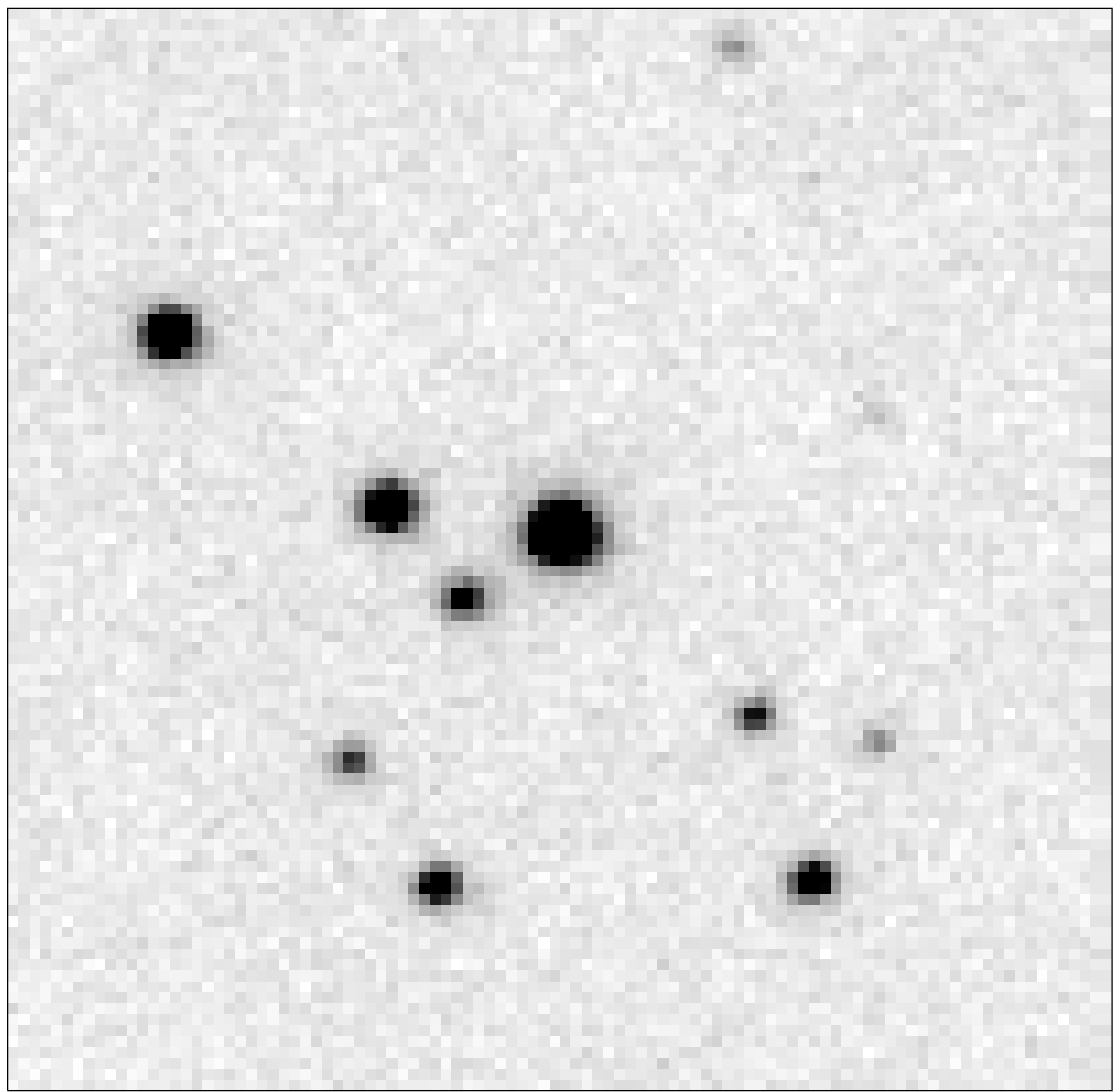}
\end{minipage}
\begin{minipage}[t!]{40mm}
 \includegraphics[width=40mm]{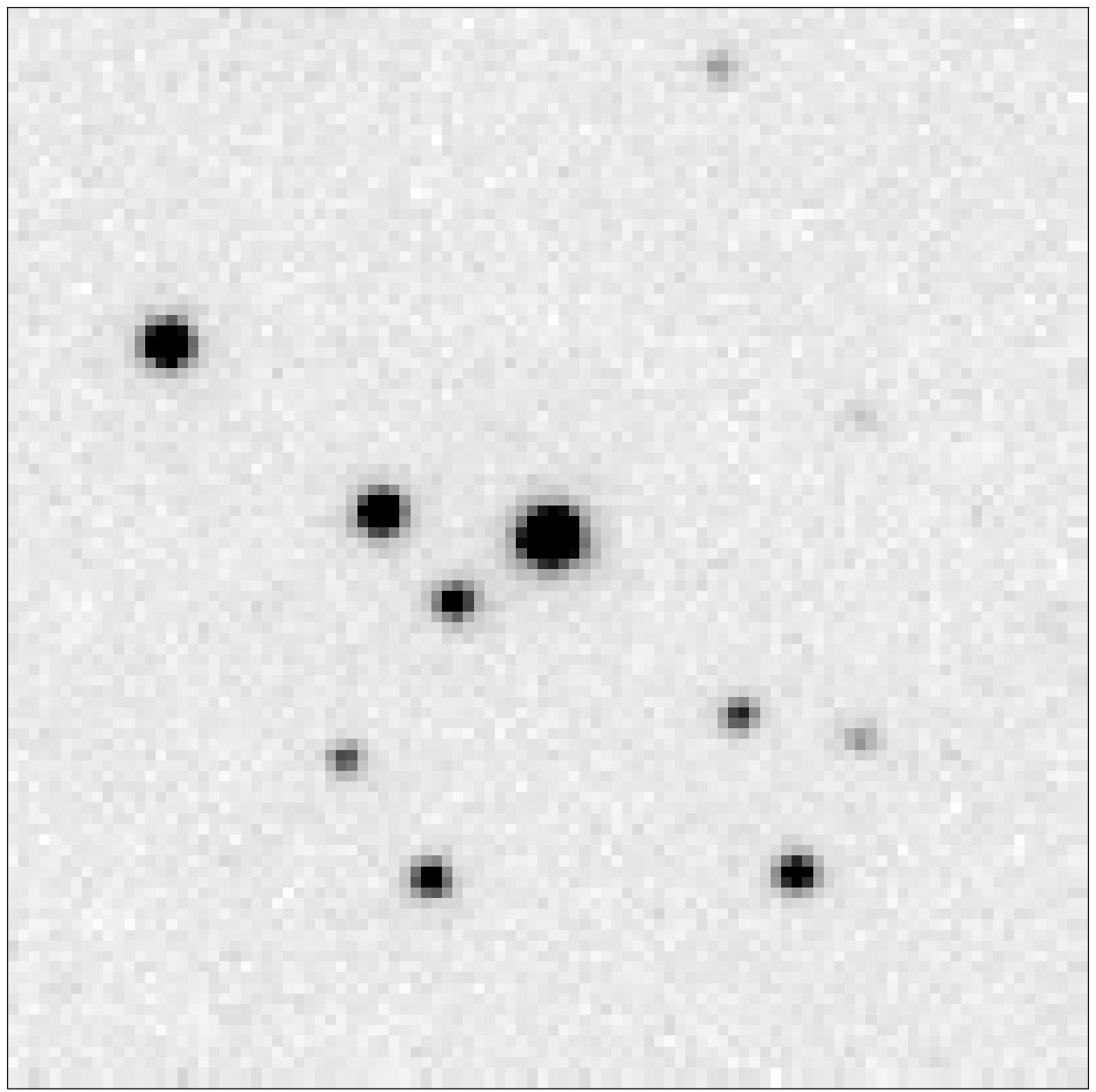}
\end{minipage}
\begin{minipage}[t!]{40mm}
 \includegraphics[width=40mm]{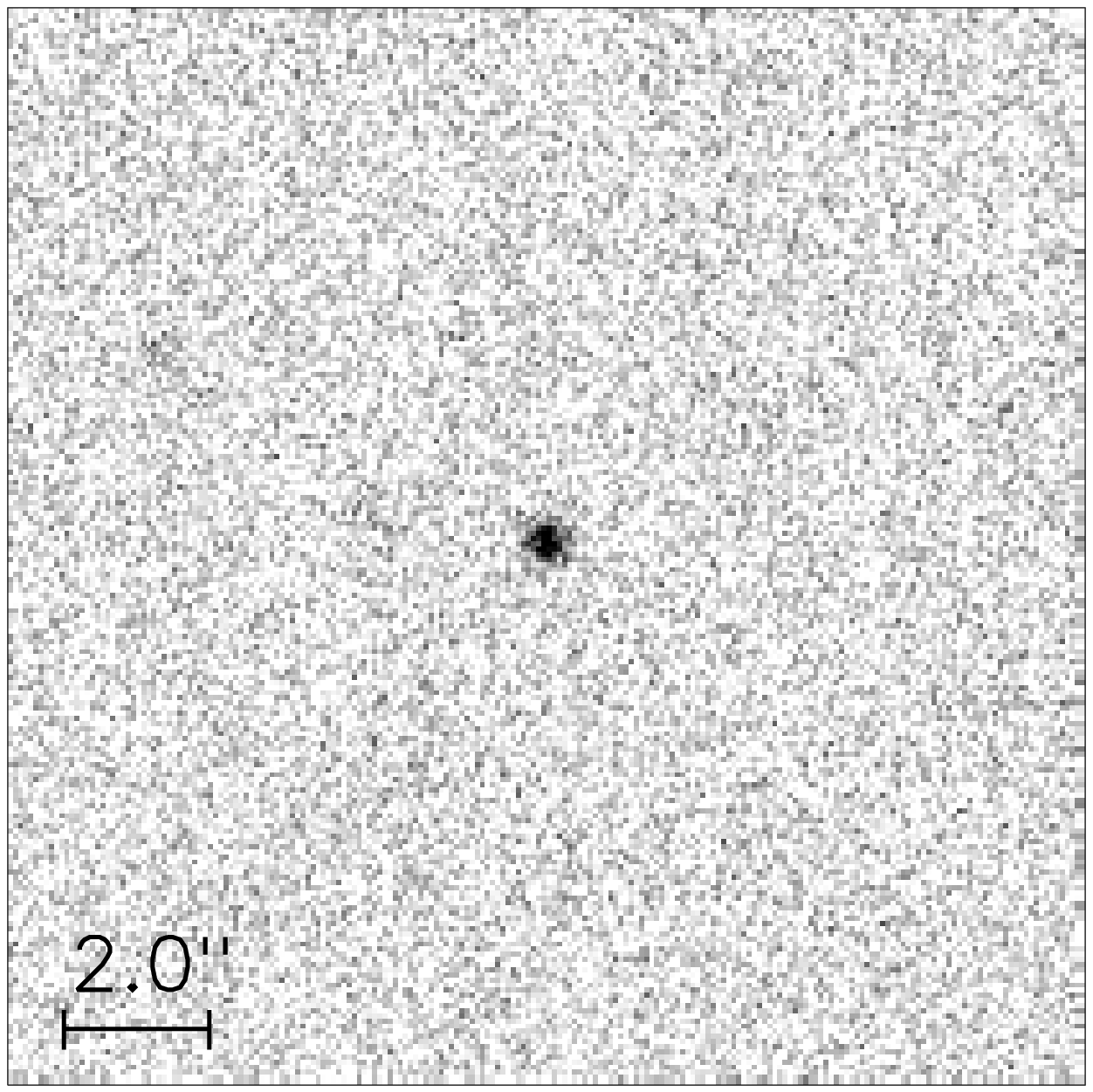}
\end{minipage}
\caption{\label{figims} Images of GALEX1931 (center star in each frame). 
               North is up and East is left in each image.
               {\it Top row:} BVRI (from left to right) images from the Nickel 1-m. All images
               are presented with the same inverted linear stretch. {\it Bottom row:} 
               J$_{\rm s}$HK$_{\rm s}$L$^{\prime}$  (from left to right) images from ISAAC. 
               The J$_{\rm s}$HK$_{\rm s}$
               images are all presented with the same inverted linear stretch. Note that the two objects 
               within $\approx$2$^{\prime\prime}$ of GALEX1931 are red.} 
\end{figure}

Nickel images are presented in Figure \ref{figims} and
photometry is reported
in Table \ref{tab1931flux} which also includes
Galaxy Evolution Explorer \citep[GALEX;][]{martin05} fluxes.




\subsection{ISAAC Near-Infrared Imaging}

Near-infrared imaging was performed
with ISAAC \citep{moorwood98} mounted on the VLT-UT3 telescope.
Data were obtained on UT 29 October 2010 at Cerro Paranal
in excellent conditions, resulting 
in stellar images with FWHM $<$0.5$^{\prime\prime}$ in the J$_{\rm s}$- and H-bands and 
$<$0.4$^{\prime\prime}$ at K$_{\rm s}$. The J$_{\rm s}$ filter was employed due to a 
red (K-band) leak in the standard J filter. Utilizing individual exposure times of 
10 s, a 9-point dither pattern with non-uniformly spaced positions inside a 
30$^{\prime\prime}$ bounding box was executed at each of J$_{\rm s}$HK$_{\rm s}$.
The standard star EG 141 \citep{hawarden01} 
was observed in an identical manner but with 5 s individual exposures.

These data were reduced at the telescope by the 
{\sf Gasgano} ISAAC pipeline. A sky frame is generated from the median of all
the science images and subtracted from individual exposures. A master flat field 
is employed to correct pixel to pixel gain variation. Individual science frames
are registered using
2D cross-correlation and averaged to yield a fully reduced image.
Photometry is performed on both GALEX1931 and the standard star with the 
IRAF task {\sf apphot} using an 8 pixel (1.2$^{\prime\prime}$) aperture 
radius and 20-30 pixel sky annulus.
Flux measurements for GALEX1931 are corrected for a mild 
airmass difference between GALEX1931 and EG 141
that amounts to $<$0.02 mag, and flux-calibrated on the 
Vega system. Although S/N$\gtrsim$300 was achieved for the science target 
at all three bandpasses, a conservative error of 5\%
is chosen for the derived magnitudes to compensate for uncharacterized 
filter transformation functions for the ISAAC filters and 
other uncharacterized systematics \citep[see e.g.,][]{leggett06}.

Imaging photometry of GALEX1931 was also performed at L$^{\prime}$ using 
both telescope nodding and secondary mirror chopping. An ABBA 
nodding pattern was repeated with several arcseconds between each nod
position and approximately 30 s spent at each nod step
(or 120 s for the entire cycle). At each nod position, chop throws of 20$^{\prime\prime}$ 
were utilized at a frequency of 0.43 Hz with individual exposure times of 0.11 s 
and 9 coadds at each mirror position. Additionally, the position 
of the ABBA pattern on the array was non-uniformly dithered between cycles within a 
10$^{\prime\prime}$ box. The total on-source time for GALEX1931 
was 1188 s, consisting of 30 chopping cycles nested within 10 nodding 
cycles of 4 positions each. The L$^{\prime}$ standard star FS 148 
\citep{leggett03} was observed similarly with a total on-source time of 
118.8 s over 1 complete nodding cycle.

A combined and reduced L$^{\prime}$ frame was generated at the telescope 
by the pipeline. The two coadded and chop-subtracted images produced 
at each consecutive pair of nod positions are combined, and then divided 
by a master flat field.  Image alignment is achieved by compensating for the 
executed telescope offsets between nodding, and the two images are averaged.
The difference in chopped images effectively removes the sky background, 
and the observing pattern results in one positive and two negative stellar
images in the recombined frame. A final, reduced image is constructed by 
combining (aligning and averaging) all the frames obtained in this manner.
Photometry on both GALEX1931 and the standard star is performed with 
{\sf apphot} using an aperture radius of 20 pixels (1.4$^{\prime\prime}$) and sky 
annulus of 30-50 pixels. The difference in airmass between GALEX1931
and FS 148 implies an extinction correction of less than 0.01 mag, and 
the flux of GALEX1931 is calibrated on the Vega system. A total photometric 
error is calculated by the quadrature sum of an 
(assumed) 5\% calibration uncertainty and the measured S/N $=9.2$ at the
photometric aperture radius.

ISAAC photometric results are listed in Table \ref{tab1931flux}. 
Figure \ref{figims} shows the ISAAC images of GALEX1931.
It is noted that the L$^{\prime}$-band flux detected from this imaging
differs significantly from the 3.35 $\mu$m flux detected toward
GALEX1931 by the WISE satellite of 1.02$\pm$0.04 mJy \citep{debes10}.
Possible reasons for this discrepancy are discussed Section \ref{secimg}.

\subsection{Optical Spectroscopy}
\label{sechobs}

Grating resolution optical spectroscopy was performed at Lick Observatory
with the KAST Double Spectrograph mounted on the Shane 3-m
telescope. Table \ref{tabhobs} lists the observation date and
instrumental setup. Data are reduced using standard
IRAF longslit tasks. Absolute flux calibration of the KAST data is
accomplished by scaling to the Nickel B- and R-band measurements and
the \citet{vennes10} V-band measurement.

Keck HIRES \citep{vogt94} optical echelle spectra were obtained for GALEX1931. 
Table \ref{tabhobs} lists the observation date and HIRES
instrumental setup. Data were reduced using the
MAKEE software package which outputs heliocentric velocity corrected spectra
shifted to vacuum wavelengths. After reduction and extraction, 
high order polynomials are fit to each order to bring overlapping 
order segments into agreement before combining both
HIRES exposures and all orders. 

\subsection{FIRE Near-Infrared Spectroscopy}

Near-infrared spectroscopy was obtained with
the Folded-port Infrared Echellete \citep[FIRE; ][]{simcoe08,simcoe10} 
mounted on the Magellan 6.5-m Baade telescope. Observation parameters
are listed in Table \ref{tabhobs}.


Prism-mode observations were performed
with a 0.6$^{\prime\prime}$ slit aligned with the parallactic angle. 
One ABBA nod-pattern was obtained for GALEX1931. A single
nod pair of 1 s integration per nod 
was obtained for the telluric calibrator star HD 189920
(A0~V). HD 189920 was too bright to be observed directly through the
slit. The telescope was offset by 6$^{\prime\prime}$ from HD 189920 and defocused
until un-saturated spectra could be obtained. Although such an observation
strategy would prevent accurate absolute flux calibration, the relative flux
calibration should not be impaired. Data reduction followed
\citet{burgasser10} with the exception of the dispersion solution being
fit with an 8th-order Legendre polynomial \citep[essentially equivalent to the
5th-order cubic spline employed by][]{burgasser10}. Absolute flux
calibration of the GALEX1931 prism data is accomplished by scaling its 
spectrum to the ISAAC J$_{\rm s}$HK$_{\rm s}$-band measurements.
The FIRE prism data are presented in Figure \ref{fig1931sed}. 

\section{Results and Modeling}

\subsection{Optical Spectroscopy}
\label{secspec}

\begin{figure}
 \centering
 \includegraphics[width=140mm]{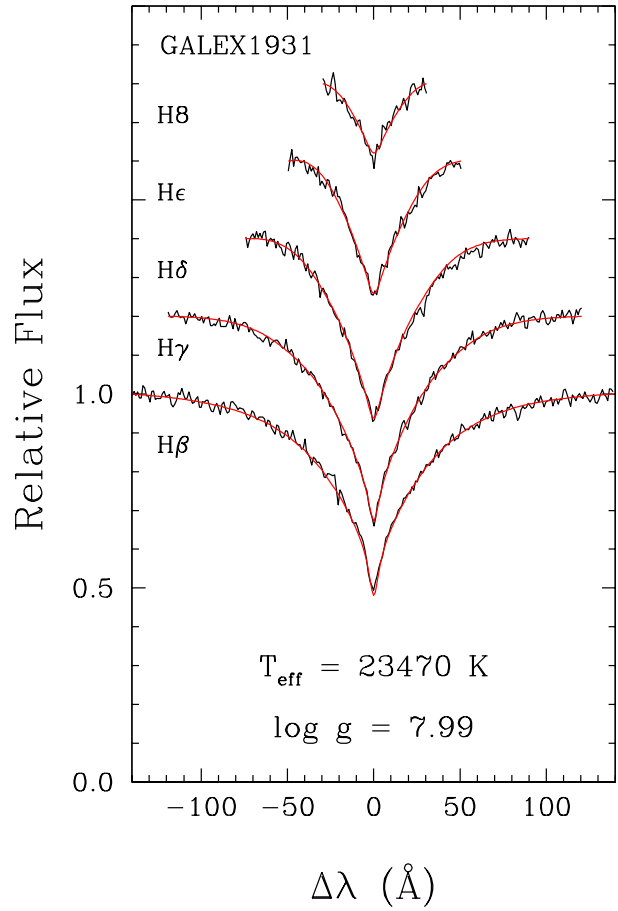}
\caption{\label{fig1931balm} Best fit models (red lines) to Balmer lines in the 
               KAST blue side data of
               GALEX1931 (underlying black spectrum). 
               The best fit T$_{\rm eff}$ and log$g$ are also displayed.}
\end{figure}

Parameters for the white dwarf star are derived from model fits to
all Balmer lines in the blue-side KAST data (Figure \ref{fig1931balm}). 
The method relies on the ``spectroscopic technique'' developed by
\citet{bergeron92} and described at length in \citet{liebert05} 
and references therein. 
We find that the KAST data are best reproduced by a white dwarf
with parameters T$_{\rm eff}$ = 23470$\pm$300 K
and log\,$g$ = 7.99$\pm$0.05 where $g$ has units of cm s$^{-2}$. 
We note that GALEX1931's effective temperature obtained
from our fitting method differs significantly from that reported
in \citet{vennes10} $-$  T$_{\rm eff}$ = 20890$\pm$120 K; the origin
of this discrepancy is not clear at this time.
By matching a model white dwarf atmosphere with our derived parameters
to the optical spectra and photometry we derive a distance
to the white dwarf of 56$\pm$3 pc $-$ this agrees within the quoted uncertainties
with the value calculated by \citet{vennes10}. It is noted that the
spectra and photometry also favor a hotter T$_{\rm eff}$ for GALEX1931
than the value reported in \citet{vennes10}. Cooling models
similar to those described in \citet{fontaine01}, except with
carbon-oxygen cores, are used to estimate a cooling age
of 30 Myr.
The white dwarf mass and envelope mass are estimated
to be M$_{\rm WD}$ = 0.63 M$_{\odot}$ and log\,$q$ = $-$16.1 
\citep[where $q$= M$_{\rm env}$/M$_{\rm WD}$;][D. Koester 2010 private communication]{koester09}.

With these parameters for GALEX1931 we proceed
in fitting the metallic absorption lines detected in the
HIRES spectrum. 
We use a local thermodynamic equillibrium (LTE) 
model atmosphere code similar to that described in 
\citet{dufour05,dufour07}. 
Absorption line data are taken from the Vienna Atomic Line Database
\begin{footnote}
\verb+http://vald.astro.univie.ac.at/~vald/php/vald.php+
\end{footnote}.
Using the effective temperature and surface gravity determined from
fitting the Balmer lines we calculate grids of
synthetic spectra for each element of interest. The grids
cover a range of abundances typically from log[$n$(Z)/$n$(H)]= $-$3.0 to
$-$7.0 in steps of 0.5 dex. We then determine the abundance of each
element by fitting the various observed lines using a similar method
to that described in \citet{dufour05}. Briefly, this is done by
minimizing the value of $\chi$$^2$ which is taken to be the sum
of the difference between the normalized observed and model fluxes over
the frequency range of interest with 
all frequency points being given an equal weight. This is
done individually for each line and the final adopted
abundances (see Table \ref{tab1931}) are taken to be the average of all the
measurements made for a given element after removing outliers. 
Uncertainties are taken to be the dispersion among abundance
values used in the average. If this dispersion is less than 0.10 dex, then
the abundance uncertainty is set to 0.10 dex.

As in \citet{vennes10}, we find contributions from Ca, Mg, Si, and Fe. The
HIRES spectra do not extend sufficiently into the red to confirm
the presence of O, so we use the \citet{vennes10} O~I $\lambda$7777 triplet equivalent
widths in place of our own measurements to estimate the O abundance.
We make synthetic spectra for the O~I 
triplet in the same way as for the other elements. The equivalent widths
of the synthetic O~I lines are then compared to the \citet{vennes10} measured
values to derive the abundance reported in Table \ref{tab1931}.
Abundance measurements for Ca, Mg, Si, and O
are in agreement with those reported in \citet{vennes10}
not withstanding the significant T$_{\rm eff}$ difference between our
white dwarf parameters and theirs. The revised abundances quoted in
\citet{vennes11} are also in agreement with our own, except for Mg which
is less abundant by $\approx$0.3 dex due to the \citet{vennes11} 
inclusion of pressure broadening
and non-local thermodynamic equilibrium effects.
However, we find
an Fe abundance higher than \citet{vennes10,vennes11} by $\approx$0.4 dex. 
It is noted that fits to the Fe lines assuming
a T$_{\rm eff}$ of 21,000 K yield an Fe abundance consistent with
that reported in \citet{vennes10}. New elements discovered in the HIRES data are
chromium identified from two Cr~II lines and
manganese identified from one Mn~II line (Figure \ref{figlines}). Aluminum
may potentially be present in the HIRES data as well (Figure \ref{figlines}),
however the S/N is marginal so we report
Al only as an upper limit (Table \ref{tab1931}).
Abundances for all elements identified to date in the atmosphere
of GALEX1931 are reported in Table \ref{tab1931} as well as upper limits
for other elements of interest (see Section \ref{secdisc}).

It is noted that the GALEX NUV and FUV measurements appear discrepant
with the white dwarf parameters derived above. We find that the model
can be brought into agreement with the data by reddening the white dwarf
with the \citet{cardelli89} extinction curve assuming E(B$-$V)= 0.03 and interstellar
reddening. However, such reddening places the model at the limit of the 3$\sigma$
uncertainties quoted
for the optical photometry. It is likely that the true color excess toward
GALEX1931 is less than 0.03 and that some combination of reddening and
metal line blanketing (from elements detected and as of yet undetected)
in the UV-spectrum of GALEX1931 are responsible for the depressed
GALEX measurements relative to the model.

\subsection{Infrared Excess Modeling}
\label{secimg}

\begin{figure}
 \includegraphics[width=160mm]{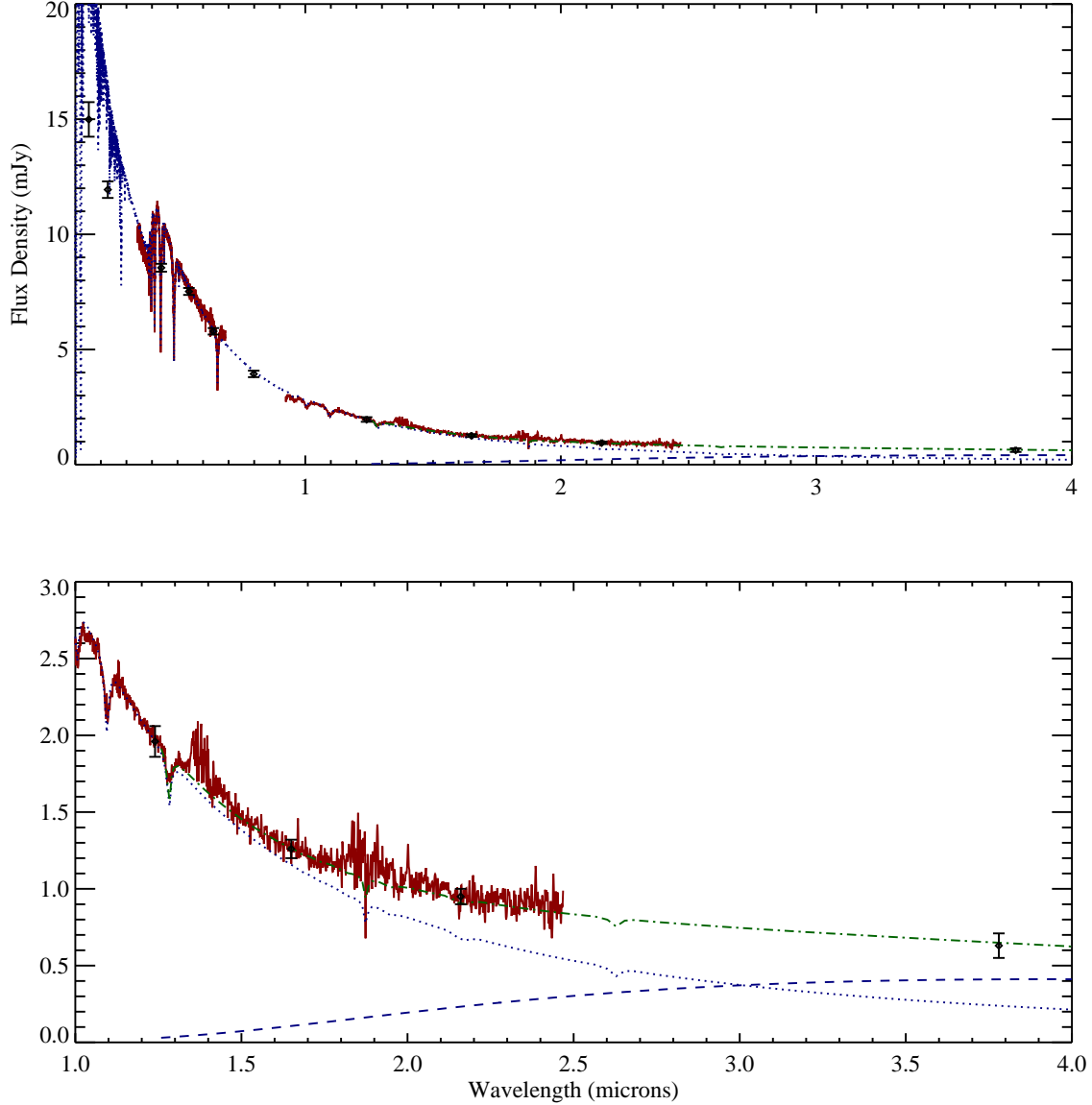}
\caption{\label{fig1931sed} Spectral energy distribution for the DAZ white dwarf GALEX1931. 
         Overplotted are models for an atmosphere of a T$_{\rm eff}$=23,500 K, 
         log($g$)=8.0, DAZ white dwarf
         with an orbiting flat, passive, opaque dust disk having
         T$_{\rm inner}$=1400 K, T$_{\rm outer}$=1200 K, and $i$=70$^{\circ}$. Data points are
         GALEX NUV and FUV (Section \ref{secspec} discusses the discrepancy between the model
         and 
         the GALEX NUV and FUV measurements), Nickel BRI, the \citet{vennes10} V, and
         ISAAC J$_{\rm s}$HK$_{\rm s}$L$^{\prime}$. The red curve in the optical
         is the KAST spectrum. The red curve in the infrared is the FIRE
         prism-mode spectrum. Both the KAST and FIRE data are flux calibrated to
         the broadband photometry. The blue, dotted curve is the white dwarf
         atmospheric model and the blue dashed curve is the disk model.
         The sum of the two models is shown with the green dash-dotted line.
         }
\end{figure}

\begin{figure}
  \includegraphics[width=130mm,angle=-90]{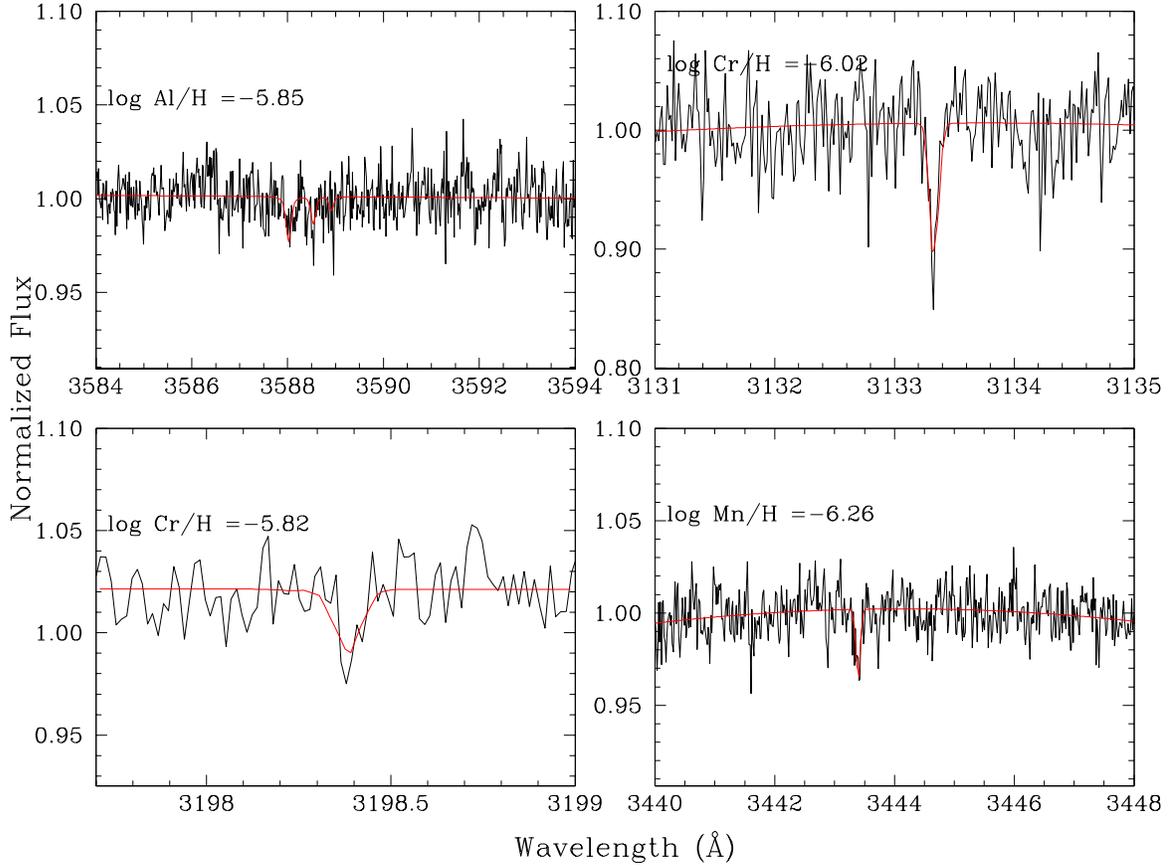}
\caption{\label{figlines} Elements identified in the HIRES spectra of GALEX1931
                not seen by \citet{vennes10}.
                Aluminum is tentatively identified, and the indicated abundance is suggested as
                an upper limit. Higher S/N spectra in this region or ultraviolet spectra could confirm
                the detection of Al. Wavelengths are in the vacuum heliocentric
                reference frame.}
\end{figure}

\begin{figure}
 \begin{center}
  \includegraphics[width=160mm]{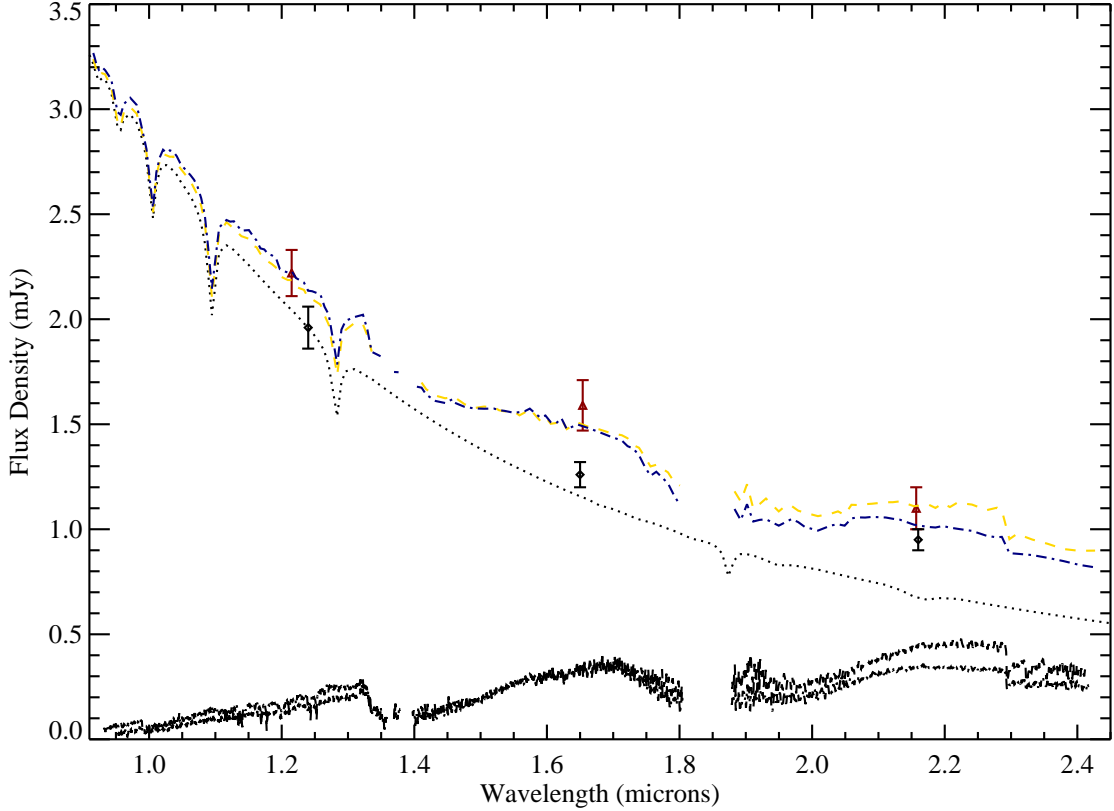}
 \end{center}
\caption{\label{fignobds} Comparison of near-infrared measurements to models
               for a roughly mid-L type brown dwarf orbiting GALEX1931. The red, upper
               data points are the 2MASS photometry; the black, lower data points are
               the ISAAC photometry. The black dotted curves at the bottom of the plot are
               mid-L type brown dwarf spectral data from the SpeX spectral library
               \citep{cushing05,rayner09} shifted in flux to the distance estimated for
               GALEX1931 (Section \ref{secspec}). The dotted curve in the middle of the
               plot is an atmospheric model for GALEX1931. The blue and gold curves
               are the sum of the white dwarf and brown dwarf curves for an L5 and L4.5
               type brown dwarf, respectively. The FIRE data smoothly connect the
               ISAAC data points.}
\end{figure}

Table \ref{tab1931flux} lists measured ultraviolet, optical, and near-infrared 
fluxes for GALEX1931. These data are presented with the KAST and FIRE prism-mode
data in Figure \ref{fig1931sed}. 
Comparison of the Table \ref{tab1931flux} J$_{\rm s}$HK$_{\rm s}$ magnitudes
to the 2MASS magnitudes
\citep[J=14.66$\pm$0.05, H=14.55$\pm$0.09, K${\rm s}$=14.45$\pm$0.10;][]{skrutskie06}
suggests that the 2MASS measurements
for GALEX1931 suffer from contamination in the photometric apertures
and/or annuli used to calculate the sky background and hence are
inaccurate. Looking at Figure \ref{figims}, one can
identify at least two sources that may have fallen within the 2MASS apertures
in this crowded field near the Galactic plane.
Similarly, the WISE data reported in \citet{debes10}
could have been contaminated by these same sources.
Due to the discrepancy between the WISE data and the high angular resolution
data presented herein, we do not include the WISE data in our analysis.
The ISAAC J$_{\rm s}$- and H-band measurements agree to
within their errors with the white dwarf photospheric model while the K$_{\rm s}$-band
measurement confirms the upward slope of the FIRE spectrum demonstrating
excess emission that begins at wavelengths near 2 $\mu$m.
This result challenges the conclusions
of \citet{vennes10} whose suggestion of an L5-type dwarf
companion relied on the H$-$K$_{\rm s}$ color
of the 2MASS photometry. Figure \ref{fignobds} illustrates 
that the excess emission implied by the contaminated 2MASS
photometry is potentially consistent with an L-type companion.
The ISAAC photometry, which is of higher fidelity, clearly
indicates that a roughly mid L-type object cannot be the cause of the
excess infrared emission. Later-type companions are ruled out by
the ISAAC photometry \citep[e.g.,][]{farihi05}
and FIRE spectra: the spectra smoothly connect the ISAAC
J$_{\rm s}$HK$_{\rm s}$ data points and show
no evidence for  the expected spectral features arising from a
sub-stellar companion \citep{farihi04}.

Given the volume of previous results linking white dwarf metal pollution
and dusty disks \citep[e.g.,][]{farihi09}, it seems reasonable to conclude
that the observed excess infrared emission detected toward GALEX1931 emanates
from a disk that orbits (and is accreted by) the white dwarf.
A model for the photospheric emission of GALEX1931 is
computed with the parameters described in Section \ref{secspec} and
including the detected metals with abundances as
listed in Table \ref{tab1931}. To fit the observed near-infrared
spectra and photometry, we add to the white dwarf model
an optically thick, flat dust disk
\citep{jura03b,jura07b}
with an inner disk temperature (T$_{\rm inner}$) of 1400 K, outer
disk temperature (T$_{\rm outer}$) of 1200 K, and an inclination angle ($i$)
of 70$^{\circ}$ (where $i$ = 90$^{\circ}$ is edge on to our line of sight). 
In this fit we set R$_{\rm WD}$/D = 5.3$\times$10$^{-12}$
(see Section \ref{secspec}). This disk and star model is shown in
Figure \ref{fig1931sed} and reproduces well the observed data. It is noted
that the fit presented here is not unique and that
the temperature of the outer region of GALEX1931's disk
is weakly constrained by the present data.

For the case of optically thick, flat dust disks Equation 1 from \citet{jura03b} can be 
used to determine at what radial distance from a white dwarf dust 
particles of certain temperatures reside. The inner disk
temperature of 1400 K  suggests particles residing as close
as 25 R$_{*}$ for a flat disk geometry. A temperature of 1400 K is close to
the sublimation temperature for some silicates, the expected mineralogical constituent of
GALEX1931's dusty disk given the stoichiometry of
other white dwarf disks \citep{jura09a,reach09} and the metals detected
in its atmosphere (Table \ref{tab1931}). Thus, the modeled inner disk radius
could be physically motivated by sublimation of grains
that will eventually be accreted by GALEX1931. Our
data set is not comprehensive enough to rule out a warped disk morphology
which could increase the inner disk radius while still reproducing the observed
near-infrared excess emission
\citep[e.g.,][and references therein]{jura07a,jura09a} .
It is noted that if a flat dust disk geometry accurately describes the
distribution of GALEX1931's orbiting debris, then the disk is likely
narrow as flat dust disk models with T$_{\rm outer}$$\leq$ 1000 K
(R$_{\rm outer}$$\leq$ 40 R$_{\rm WD}$) are not consistent
with the data. Such a result suggests GALEX1931 may host a 
narrow distribution of disk material \citep[e.g., see][]{farihi10a}. Observations at longer
wavelengths should solidify which of the above interpretations
are valid.

In their red echelle spectra of GALEX1931,
\citet{vennes10} do not identify double-peaked
emission lines from the Ca~II infrared triplet indicative of an orbiting gaseous
disk \citep{gaensicke06,melis10}.
Modeling and results presented by \citet{melis10} and
\citet{jura08} suggest that gas disks around white dwarfs are fed
by an influx of rocky objects into the Roche limit of a white dwarf star. 
The absence of such emission lines and hence an obvious gas disk orbiting 
GALEX1931 are suggestive of its dust disk originating in the tidal
shredding of one large rocky object.

\section{Discussion}
\label{secdisc}

Modeling of the ultraviolet through infrared spectral energy distribution
of GALEX1931 provides strong evidence that an orbiting dusty circumstellar 
disk emits the detected excess infrared emission. Following
earlier results, we assume that this dusty disk has its origin in the tidal
disruption of a rocky object that breached GALEX1931's Roche limit.
The ultimate fate of such material is to be accreted by the white dwarf star.
Under this assumption we can deduce the bulk elemental composition
of such an object from the metal abundances reported in Table \ref{tab1931}.
To properly convert the measured abundances to the abundances
at the time of accretion we must include the effects of diffusion,
radiative levitation, and the relevant accretion phase
\citep[][and references therein]{koester09,chayer10,jura09b}. The
presence of a disk orbiting GALEX1931 and the extremely
short elemental diffusion constants (Table \ref{tab1931}) imply that we are observing
elemental abundances during the steady-state accretion phase.
In this phase, accretion of metals has reached an equilibrium
with the rate at which these metals diffuse out of the white dwarf's
photosphere. The work of \citet{chayer10} suggests that for accretion
rates $\gtrsim$10$^4$ g s$^{-1}$ the effects of radiative
levitation are negligible $-$ all measured accretion rates and
upper limits exceed this threshold value (Table \ref{tab1931}) and hence we ignore
radiative levitation. Thus, we determine the accreted
abundance ratios by employing Equation 16 from \citet{jura09b}
and the diffusion constants of \citet{koester09} computed specifically
for the GALEX1931 parameters determined in Section \ref{secspec} 
(D. Koester 2010, private communication). 
Accreted abundances relative to Fe are reported in Table \ref{tab1931}.

We desire to make a rough estimate of the mass accreted to date onto
GALEX1931. Accretion rates for each element are computed under
the assumption that material accretes in the steady-state; these
values are reported in Table \ref{tab1931}. The total accretion rate
summed over all elements is $\sim$3.9 $\times$ 10$^9$ g s$^{-1}$.
An estimate of how long GALEX1931 has been accreting from
its dusty disk comes from the percentage of DA stars with effective
temperatures similar to GALEX1931 and
atmospheric pollution. It is assumed that this
percentage is linked to the length of time in which such DA white dwarf stars
would show metal pollution and hence be polluted from
a surrounding disk of material. From \citet{koester05a} and \citet{koester09b}
we find that 5 out of 116 or 4.31\% of DA white dwarfs within the
temperature range of 19,500 K $<$ T$_{\rm eff}$ $<$ 25,000 K are
polluted with metals (including GALEX1931). Taking the cooling time
of a white dwarf in the above T$_{\rm eff}$ range to be equal
to that of GALEX1931 (30 Myr, see Section \ref{secspec}), 
and assuming all white dwarfs accrete from a disk during this interval, it is
determined that the disk lifetime for GALEX1931 is 1.3 Myr.
Assuming GALEX1931 has been accreting at a constant rate for $\sim$1 Myr, we
estimate that $\sim$2 $\times$ 10$^{23}$ g of material has
accreted onto the white dwarf star. This is
close to the second most massive asteroid in
our Solar System, Vesta ($\approx$3 $\times$ 10$^{23}$ g). It is noted
that numerous assumptions are included in this estimate and that
it could increase or decrease significantly and that the estimate
does not include the mass remaining in the disk.

Following \citet{klein10}, we attempt to balance the accreted oxygen
abundances under the expectation that all oxygen was contained
in rocky oxygenated minerals (e.g., MgO, Al$_2$O$_3$, SiO$_2$, CaO, TiO$_2$, Cr$_2$O$_3$,
MnO, FeO, Fe$_2$O$_3$, and NiO) and 
water \citep[H$_2$O; see][for a discussion of water survival in rocky objects during post-main sequence evolution]{jura10}. 
The criterion that needs to be met is Equation 3 of \citet{klein10}:

\begin{equation}
 O_{bal} \equiv \sum_{Z} \, \frac{q(Z)}{p(Z)} \, \frac{n(Z)}{n({\rm O})} = 1 \, ,
\end{equation}

\noindent where element $Z$ is contained in molecule
$Z_{q(Z)}$O$_{p(Z)}$. In this equation one must consider only
hydrogen accreted from the parent body ($H_{acc}$). 
We find that the oxygen accreting
onto GALEX1931 can be accounted for almost entirely by inclusion in rocky
minerals. There is, however, the ambiguity of how Fe was partitioned in
the parent body that accreted onto GALEX1931. In our model 
Fe could have been contained in any one of metallic Fe, FeO, or Fe$_2$O$_3$
\citep[see discussion in][]{klein10}. Depending on how we
divide Fe into these three states, there can be anything from a significant deficit in the
oxygen balance ($O_{bal}$ = 0.58 instead of 1.0, indicating
too much oxygen for the number of heavier element atoms) 
to a significant overestimate of rocky
minerals ($O_{bal}$ = 1.33 instead of 1.0, indicating too little oxygen
for the number of heavier element atoms). Assuming as did \citet{klein10}
that half of the Fe is in the form of metallic Fe and half in the form of FeO, we
arrive at a value of $O_{bal}$=0.83. With this value we constrain the amount
of water present in the object that accreted onto GALEX1931 to be 
$\frac{1}{2} \, \frac{n(H_{acc})}{n(O)}$$<$0.2 or $<$1\% of the mass of
the parent body.
It is noted that use of the
lower Mg abundance derived by \citet{vennes11} would result in a more
water-rich body.
Thus it appears as though the material accreting onto GALEX1931
is very likely to have originated in rocky minerals and is likely to be only
a small percentage of water by mass.

\begin{figure}
 \begin{center}
  \includegraphics[width=160mm]{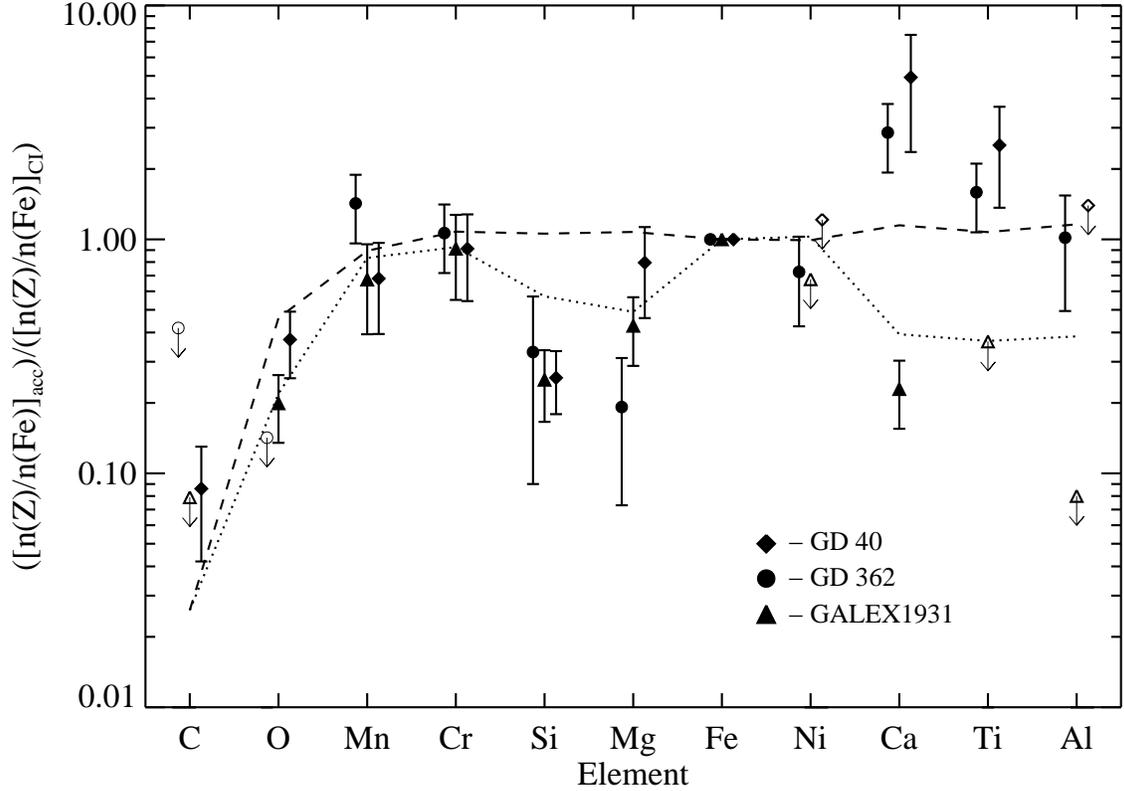}
 \end{center}
\caption{\label{figwdpolfe} Comparison of white dwarf abundances relative
               to Fe to CI Chondrite abundances relative to Fe \citep[after Figure 15 of ][]{klein10}.
               Elements have increasing condensation temperature to the right.
               The triangle data points are measurements and upper limits for GALEX1931
               from Table \ref{tab1931}. The diamond data points are measurements and
               upper limits for GD 40 \citep{klein10} while the circle data points are measurements
               and upper limits for GD 362 \citep{zuckerman07,koester09}. The dashed line
               is data for the bulk Earth \citep{allegre95,vanthienen07}. The dotted line is
               a ``wind-stripped Earth'' model as described in Section \ref{secdisc}.}
\end{figure}

Examination of the bulk elemental composition of the material that
is accreted by GALEX1931 can potentially yield insight into the evolutionary
processes that affect terrestrial-like planets during the late stages
of stellar evolution. In Figure \ref{figwdpolfe} we plot the elemental
abundances relative to Fe for three white dwarfs: GD 362
\citep{zuckerman07,koester09}, GD 40 \citep{klein10}, and
GALEX1931. We normalize each of these relative abundance
measurements (at the time of accretion for each object) to
elemental abundances relative to Fe for CI Chondrites
\citep[see Table \ref{tab1931} and][]{lodders03}. CI Chondrites
are chosen as these objects have some of the best-determined
bulk compositions of any object in our Solar system. CI Chondrites
have near-Solar compositions and thus represent primitive rocks
that likely formed directly from the Solar nebula \citep[e.g.,][and references therein]{lodders03}.
Under the assumption that primitive rocks are similar in all
planetary systems, we seek to probe whether or not the parent body being
accreted by GALEX1931 has undergone significant processing
by comparing the GALEX1931 measured abundances to CI Chondrite
abundances. To further place
the white dwarf measurements into context, we overplot the bulk
elemental composition of Earth as extracted from the
measurements of \citet{allegre95} and the compilation
of \citet{vanthienen07}. 

Most striking in Figure \ref{figwdpolfe} is the rough agreement
of the abundances of all elements for the three white dwarfs
with the bulk Earth, especially the depletion of the volatiles C and O relative
to what would be expected from the more primitive CI Chondrites. 
Next is the good agreement between transition metal abundances
(Mn, Cr, Fe, and Ni) for the accreted objects and the bulk Earth. The
obvious discrepancies are with Si and Mg and
the refractories Ca and Ti. 
Assuming the parent bodies that are accreted by each white dwarf
started close to Earth-like in bulk elemental composition
(a risky assumption, but the best that can be made at present), 
significant deficits of Si and varying
degrees of deficient Mg suggest crustal and mantle loss. This
result suggests that each white dwarf has accreted objects massive
enough to experience differentiation. That is, the atoms layered
themselves in the parent body by mass (with the heaviest atoms
at the parent body's core). Such a result indicates that the parent
bodies now being accreted by each white dwarf were indeed
massive bodies and perhaps even full-fledged planets. 

Ca, Ti, and Al abundances could potentially bear the signature
of a particular crust and mantle loss mechanism. \citet{klein10}
compared the accreted abundances for GD 40 to a model
developed by \citet{fegley87} that described the evolution of
a rocky planetary body when it is exposed to intense radiation.
The predictions of this model are in good agreement with the
parent body abundances for the object accreted by GD 40
\citep{klein10}. Such a model could potentially provide a reasonable
explanation for the parent body composition of the object accreted
by GD 362 as it too shows the characteristic enhancement of Ca and Ti
predicted in the \citet{fegley87} model, 
although the mantle loss appears to be more severe
as per the significant Mg deficiency (Figure \ref{figwdpolfe}).
To explain the results for GALEX1931 requires either additional
evolutionary processes or a different evolutionary path that
results in crust and mantle loss without the enhancement of
Ca and Ti. Indeed, GALEX1931 shows a significant
$deficit$ of Ca, Ti, and Al relative to the Earth and
CI Chondrites (see Figure \ref{figwdpolfe}). 

To explain the distinctive elemental mixture accreting onto 
GALEX1931 we propose a scenario in which the dominant
evolutionary mechanism acting on the parent body was
interaction with the host star stellar wind. After irradiation
by the luminous host star while on the asymptoptic giant branch
(AGB), interaction with the AGB ejecta seems a likely choice for
a rocky planet mass loss mechanism \citep[e.g.,][]{jura08}. 
In such a scenario
the rocky planet is essentially sanded down by the AGB wind,
losing first its crust then mantle depending on the severity of the
planet's interaction with the stellar wind. We estimate to first order
the relative abundances of such an object after crustal
and mantle stripping. We begin with the assumption that the object
started with the same composition as the bulk Earth and the same crust and
mantle abundances as the Earth \citep{anderson89,allegre95}. 
Also inherent is the assumption
that the parent body's tidally shredded remnants are well mixed in the disk.
We remove varying fractions of the crust, then upper mantle, and
then lower mantle. After this removal step, the bulk abundances
of the object are recomputed relative to Fe. Experimentation with
varying degrees of crust and mantle removal suggests that
the best match between the simple stripping model and
the GALEX1931 abundances shown in Figure \ref{figwdpolfe}
is obtained after stripping the entire crust and upper mantle
and half of the lower mantle.
The simple model provides good agreement with the measurements,
with the exception of aluminum. 
These results suggest that ``wind stripping'' of the outer layers of an
Earth-like planet is a viable mode to produce the observed parent body
abundances in at least some white dwarfs and thus that
interaction with the AGB stellar wind is likely an important process in a
terrestrial planet's final evolutionary phase. Exploration
of the relevant physics in the wind stripping process can further develop
the simple model laid out above.

\section{Conclusions}

We have carried out a comprehensive suite of observations aimed
at characterizing GALEX1931's infrared excess and atmospheric
pollution. Our broad data set allows us
to identify the excess infrared emission detected toward GALEX1931
as emanating from an orbiting dusty debris disk. Analysis of the heavy
metals detected in GALEX1931's atmosphere suggests that
it is accreting the remnants of a differentiated $-$ and hence
massive $-$ rocky body. Assuming the object being accreted by
GALEX1931 started out roughly Earth-like in composition,
comparison of the distinctive bulk elemental composition of this
terrestrial-like planet to predictions from a simple wind-stripping model 
suggests that the planet's outer
layers were removed before the planet was accreted by the white dwarf
star. A likely culprit for such stripping is interaction between the planet and
its host star's dense stellar wind while the star progresses through the
asymptoptic giant branch.



\acknowledgments

C.M.\ was supported by the National Science Foundation under award No.\
AST-1003318. 
P.D. is a CRAQ postdoctoral fellow.
We thank Detlev Koester for kindly providing diffusion constants
tailored to GALEX1931.
Some of the data presented herein were obtained at the W.M. Keck Observatory, 
which is operated as a scientific partnership among the California Institute of 
Technology, the University of California and the National Aeronautics and Space 
Administration. The Observatory was made possible by the generous financial 
support of the W.M. Keck Foundation.
This paper includes data gathered with the 6.5 meter Magellan 
Telescopes located at Las Campanas Observatory, Chile.
This publication makes use of data products from the Two Micron All Sky
Survey, which is a joint project of the University of Massachusetts and the
Infrared Processing and Analysis Center/California Institute of Technology,
funded by the National Aeronautics and Space Administration and the National
Science Foundation. This research has made use of the SIMBAD database.
Based on observations made with the NASA 
Galaxy Evolution Explorer. GALEX is operated for NASA by the California 
Institute of Technology under NASA contract NAS5-98034.
This research was supported in part by NASA grants to UCLA.
This work is supported in part by the NSERC Canada and by the Fund FQRNT (Qu\'ebec).



{\it Facilities:} \facility{Keck:I (HIRES)}, \facility{VLT:Melipal (ISAAC)}, \facility{Magellan:Baade (FIRE)}, \facility{Nickel (Direct Imaging Camera)}, \facility{Shane (KAST)}




\clearpage

\clearpage

\clearpage

\clearpage

\clearpage

\clearpage

\clearpage

\begin{deluxetable}{cccc}
\tabletypesize{\normalsize}
\tablecolumns{4}
\tablewidth{0pt}
\tablecaption{Broad-band Fluxes for GALEX1931 \label{tab1931flux}}
\tablehead{ \colhead{Band} & \colhead{$\lambda$} & \colhead{$m$\tablenotemark{a}} & \colhead{F$_{\rm obs}$} \\
                      \colhead{}           & \colhead{(nm)}              & \colhead{(mag)}         & \colhead{(mJy)}
}
\startdata
\cutinhead{ISAAC$-$ NIR\tablenotemark{a}} \\
L$^{\prime}$ & 3780 & 14.00$\pm$0.14 & 0.63$\pm$0.08\tablenotemark{b} \\
K$_{\rm s}$ & 2160  & 14.62$\pm$0.05 & 0.95$\pm$0.05  \\
H                   & 1650  & 14.80$\pm$0.05 & 1.26$\pm$0.06   \\
J$_{\rm s}$  & 1240  & 14.80$\pm$0.05 & 1.96$\pm$0.10 \\
\cutinhead{Nickel 40$^{\prime\prime}$\tablenotemark{a} $-$ Optical} 
 I  &  798.2    & 14.47$\pm$0.04 & 3.94$\pm$0.16 \\
 R &  640.7   & 14.31$\pm$0.03 & 5.79$\pm$0.17 \\
 V &  544.8    & 14.20$\pm$0.02 & 7.52$\pm$0.15 \\
 B &  436.3   & 14.18$\pm$0.02 & 8.55$\pm$0.17 \\
\cutinhead{GALEX\tablenotemark{a} $-$ Ultraviolet} 
NUV  &     227.1   &  13.70$\pm$0.03 &   11.94$\pm$0.36 \\
FUV  &     152.8   &  13.46$\pm$0.05  &   14.99$\pm$0.75 \\
\enddata
\tablenotetext{a}{ISAAC and Nickel magnitudes are on the Vega system.
GALEX measurements are in AB magnitudes and have
uncertainties as suggested in \citet{morrissey07}.}
\tablenotetext{b}{This value disagrees with the WISE 3.35 $\mu$m measured value
of 1.02$\pm$0.04 mJy \citep{debes10}, see Section \ref{secimg}.}
\end{deluxetable}

\clearpage

\begin{deluxetable}{lccccccc}
\rotate
\tabletypesize{\normalsize}
\tablecolumns{8}
\tablewidth{0pt}
\tablecaption{Spectroscopic Observations Summary \label{tabhobs}}
\tablehead{
\colhead{UT Date} & \colhead{Instrument} & \colhead{Setup} & \colhead{Coverage} & \colhead{Resolution\tablenotemark{a}} & \colhead{Integration Time (s)} & \colhead{S/N} 
 & \colhead{$\lambda$ of S/N\tablenotemark{b} (\AA )}
 }
\startdata
05 July 2010 & HIRES & UV Collimator  & 3130-5940 \AA\ & $\sim$40,000 & 2$\times$1800 & 80\tablenotemark{c} & 3450 \\
19 September 2010 & FIRE & Prism-mode & 0.7-2.5 $\mu$m & $\approx$250-350\tablenotemark{d} & 4$\times$60 & 80\tablenotemark{c} & 10500 \\
\multirow{2}{*}{09 October 2010\tablenotemark{e}} & \multirow{2}{*}{KAST} & 600/4310  & 3430-5510 \AA\ & $\sim$3.4 \AA\ & 600 & 100 & 4500 \\
                                                                &            & 1200/5000 & 5520-6910 \AA\ & $\sim$2.4 \AA\ & 600 & 60 & 6300 \\
\enddata
\tablenotetext{a}{Resolutions were measured from the FWHM of single arclines in our comparison spectra.}
\tablenotetext{b}{Wavelength where S/N measurement was made in the spectrum.}
\tablenotetext{c}{S/N for combined exposures.}
\tablenotetext{d}{See \citet{burgasser10}.}
\tablenotetext{e}{Observations used the 2.0$^{\prime\prime}$ slit and the d55 dichroic.}
\end{deluxetable}

\clearpage

\begin{deluxetable}{cccccc}
\rotate
\tabletypesize{\normalsize}
\tablecolumns{6}
\tablewidth{0pt}
\tablecaption{GALEX1931+0117 Metal Pollution \label{tab1931}}
\tablehead{ 
  \colhead{$Z$} & 
  \colhead{log[$n$($Z$)/$n$(H)]$_{\rm measured}$} & 
  \colhead{$\tau$$_{\rm diff}$ (days)\tablenotemark{a}} &
  \colhead{[$n$($Z$)/$n$(Fe)]$_{\rm accreted}$\tablenotemark{b}} &
  \colhead{[$n$($Z$)/$n$(Fe)]$_{\rm CI}$\tablenotemark{c}} &
  \colhead{$\dot{M}$$_{\rm acc}$/(10$^7$ g s$^{-1}$)\tablenotemark{d}}
}
\startdata
C    & $<$$-$4.85 & 5.811 & $<$0.0712 & 0.8912 &  $<$3.15 \\ 
O & $-$3.68$\pm$0.10\tablenotemark{e} & 3.534 & 1.7338 & 8.7096 & 102.0  \\ 
Mg  &  $-$4.10$\pm$0.10 & 4.532 & 0.5140 & 1.2022 & 46.1 \\ 
Al    & $<$$-$5.85 & 5.336 & $<$0.0077 & 0.0954 & $<$0.772 \\ 
Si    & $-$4.35$\pm$0.11 & 4.542 & 0.2884 & 1.1481 & 29.8 \\  
Ca  & $-$5.83$\pm$0.10 & 2.730 & 0.0158 & 0.0691 & 2.34 \\ 
Ti    & $<$$-$7.00 & 2.960 & $<$0.0010 & 0.0027 & $<$0.174 \\ 
Cr   & $-$5.92$\pm$0.14 & 2.560 & 0.0137 & 0.0151 & 2.55 \\ 
Mn  & $-$6.26$\pm$0.15 & 2.292 & 0.0070  & 0.0104 &  2.64 \\ 
Fe   &  $-$4.10$\pm$0.10 & 2.329 & 1.00 & 1.00 & 205.0 \\ 
Ni    & $<$$-$5.60 & 1.982 & $<$0.0371 & 0.0549 & $<$8.04 \\ 
\enddata
\tablenotetext{a}{Diffusion constants, see \citet{koester09} and Section \ref{secdisc}.}
\tablenotetext{b}{Parent body abundances relative to Fe; see Section \ref{secdisc} and Figure \ref{figwdpolfe}.}
\tablenotetext{c}{CI chondrite data from \citet{lodders03}. See also Figure \ref{figwdpolfe}.}
\tablenotetext{d}{$\dot{M}$$_{\rm acc}$($Z$)=$M_{\rm env}$($Z$)/$\tau$$_{\rm diff}$($Z$) where $M_{\rm env}$($Z$) is the mass of element $Z$ in GALEX1931's envelope assuming the hydrogen-dominated envelope mass is 9.4 $\times$ 10$^{16}$ g \citep[][D. Koester 2010 private communication]{koester09}.}
\tablenotetext{e}{This value is derived from the equivalent widths reported in \citet{vennes10}, see Section \ref{secspec}.}
\end{deluxetable}



\end{document}